\documentclass[aps,preprint,nofootinbib,floatfix]{revtex4}
\usepackage{graphicx}
\usepackage[labelfont=bf,labelsep=quad,justification=justified]{caption}
\usepackage[latin1]{inputenc}
\usepackage{amsmath,amssymb}
\usepackage{latexsym,amssymb} 
\usepackage{hyperref}
\usepackage{epstopdf}
\usepackage{graphicx} 
\usepackage[english]{babel}
\usepackage{tabularx}
\usepackage{listings} 
\usepackage{enumerate} 
\usepackage{fancyhdr} 
\usepackage{pstricks} 
\usepackage{color}
\usepackage{subfig} 
\usepackage{array,multirow} 
\usepackage{slashed} 
\usepackage{verbatim} 
\usepackage{feynmf}
\newcommand{\lsim}{\mathrel{\mathop{\kern 0pt \rlap
  {\raise.2ex\hbox{$<$}}}
  \lower.9ex\hbox{\kern-.190em $\sim$}}}
\newcommand{\gsim}{\mathrel{\mathop{\kern 0pt \rlap
  {\raise.2ex\hbox{$>$}}}
  \lower.9ex\hbox{\kern-.190em $\sim$}}}

\def  \bcen   {\begin{center}}
\def  \ecen   {\end{center}}
\def  \beq    {\begin{equation}}
\def  \eeq    {\end{equation}}
\def  \beqa   {\begin{eqnarray}}
\def  \eeqa   {\end{eqnarray}}

\def\bea{\begin{eqnarray}}
\def\eea{\end{eqnarray}}

\begin{document}
\title{A Model with New Signatures of Exotic Chiral Fermions Coupling with Scalar Leptoquarks}
\author{
 D. A. Camargo\footnote{dacamargov@gmail.com}
}

\affiliation{
Centro de Ciencias Naturais e Humanas, Universidade Federal do ABC,
Avenida dos Estados, 5001, Santo Andr$\acute{e}$-SP, Brazil
}
\date{\today}

\begin{abstract}

In this paper, the possibility to have new physics is considered at the scale of the electroweak symmetry breaking (EWSB) in a completely new framework.  
A new non-standard minimal chiral fermion extension of the Standard Model (SM) which renders the complete model free from anomalies is included.
The new fermionic matter coupled with scalar leptoquaks (SLQ) developing new signals for exotic fermions that could be evaluated 
in the Large Hadron Collider (LHC). Analysis based on the cross-section final states shows that the channels $D_{-1/3}\overline{D}_{-1/3}\rightarrow jj+\notin$ and 
$D_{-1/3}\overline{D}_{-1/3}\rightarrow 2l^{-}2l^{+}+jj$ have an apparent narrow window $M_{D_{-1/3}}\sim 240- 300$ GeV still allowed by  
direct and indirect constraints and then interesting for analysis to the new run of the LHC.

\end{abstract}
\maketitle

\section{Introduction. \label{section:1}}
The Standard Model (SM) has achieved great success in the last decades. After the discovery of the Higgs boson \cite{atlas-cms-higgs}, 
the main focus of the LHC experiments
is to discover sings of new physics. Searches based on more detailed models, like supersymmetry, in their decay chains that lead to  
easy-to-see signatures have not exhibited any particular deviation from the SM predictions. The LHC is ending the technical pause and has restarted 
a new run at a higher center of mass energy close to 14 TeV. The new run will allow the experiments to access higher mass scales, but also 
processes that gave too flimsy cross-sections to be detected during the 8 TeV run.\\
The focus of this paper is on one of such cases, where the new fields are new chiral fermions that mixes with scalar leptoquaks producing 
new signatures and decay chains capable of being detected by the LHC. The new interactions of the exotic quarks contrast with models 
containing this kind of chiral fermions, in which the decays are mediated via electroweak interactions with SM fermions 
\cite{Cacciapaglia:2011fx}-\cite{Alves:2013dga}. Therefore, the new framework with SLQ will provide new channel searches for the exotic fermions
considered here. Concerning to Leptoquaks, they are highly motivated by more elaborate theories and are expected to 
exist in various extensions of the SM such as in the Pati-Salam model \cite{Pati:1974yy}, grand unification theories based on SU(5) 
\cite{Georgi:1974sy} and SO(10) \cite{Fritzsch:1974nn} and lately, they offered the possibility to explain one of the last experimental
deviations at the LHC \cite{Allanach:2015ria}.\\
This paper presents the prospects and exclusions limits for a particular extension of the SM which contains a minimal set  
(into the $SU(2)_{L}\otimes U(1)_{\mathtt{Y}}$ gauge group) of exotic chiral fermions mixing 
with SLQ at the energies of the EWSB. The new chiral fermions form nontrivial multiplets of the SM symmetry group
$SU(2)_{L}\otimes U(1)_{\mathtt{Y}}$ of the electroweak interactions. In addition, a second scalar doublet is introduced in order to 
give mass for the exotic fields and together with a discrete symmetry $Z_{4}$ keep under control the SM predictions.

\section{Generalities of the Model. \label{section:2}}

The complete model is parametrized through an effective Lagrangian $\mathcal{L}_{Eff}$ as follows,

\begin{equation}
\begin{gathered}
\mathcal{L}_{Eff}=\mathcal{L}_{SM}+\mathcal{L}_{NF}+\mathcal{L}_{LQ-Q}
\end{gathered}
\end{equation}

where $\mathcal{L}_{SM}$ represents the SM-Lagrangian, $\mathcal{L}_{NF}$ the Lagrangian of the new field (Exotic and SLQ) and 
$\mathcal{L}_{LQ-Q}$ contain all the interactions among exotic fermions, SM-fermions and SLQ. The minimal non-standard chiral fermion extension 
is for the quark sector:

\begin{eqnarray}
&  &\cr
 &  & \psi_{L}^{Y}\equiv\left
 [\begin{array}{c}
D_{L}\\
Y_{L}
\end{array}
\right]\sim\left(\mathbf{3,2,\,}-5/6\right),\quad D_{R}\sim\left(\mathbf{3,\,1,\,}-1/3\right),\quad Y_{R}\sim\left(\mathbf{3,1,\,}-4/3\right),
\end{eqnarray}

and for the lepton sector,

\begin{eqnarray}
&  &\cr
 &  & \xi_{L}^{Y}\equiv\left
 [\begin{array}{c}
E^{3e}_{L}\\
E^{2e}_{L}
\end{array}
\right]\sim\left(\mathbf{1,2,\,}5/2\right),\quad E^{3e}_{R}\sim\left(\mathbf{1,\,1,\,}3\right),\quad E^{2e}_{R}\sim\left(\mathbf{1,1,\,}2\right),
\end{eqnarray}

In addition, two scalar leptoquarks are introduced in order to offer new possible channels for the decays of the exotic quarks. 

\begin{equation}
\begin{gathered}
\tilde{S_{u}}\sim (\mathbf{\overline{3},1},-2/3)\\
S_{1/2}\sim (\mathbf{3,2},1/6)
\end{gathered}
\end{equation}
Due to the chiral structure of the new fermionic matter, bare mass terms common in models with vector-like fermions are absent here. 
Therefore, we consider two Higgs doublet for breaking the electroweak symmetry, $\phi_{1,2}\sim (2,1)$ \cite{Bernon:2015qea} 
-\cite{Branco:2011iw}, with vacuum expectation values $\langle \phi_{1,2}\rangle=[0\hspace{0.2cm}v_{1,2}/\sqrt{2}]^{T}$. The first Higgs doublet, 
$\Phi_{1}$, is assumed to have a tree level couplings to the SM quarks fields, while the second one, $\Phi_{2}$, with the exotic
quarks fields as is done in \cite{Alves:2013dga}. For supporting this we take into account the existence of a discrete symmetry
$Z_{4}$ remaining from some physics at very high energy scale, such that the fields transforming
non-trivially have charges as shown in table \eqref{table1}. In addition, the model is assumed  into an 
special scenario of the two Higgs doublet model which entails to have a benchmark point in which the Higgses decoupled.
In this setting, the mixing angle $\alpha$ in \eqref{Higgs} takes a zero value forbidden in this way tree-level interactions
between the exotic and SM fermions and thereby protects the production of the SM-Higgs through gluon fusion.

The Yukawa Lagrangian invariant under $SU(3)\otimes SU(2)\otimes U(1)\otimes Z_{4}$ reads,

\begin{eqnarray}
-\mathcal{L}_{Y}^{Q} & = & y_{ij}^{u}\overline{q_{iL}}\widetilde{\Phi}_{1}u_{jR}
+y_{ij}^{d}\overline{q_{iL}}\Phi_{1}d_{jR}^{\prime}
\nonumber \\
 & + & y^{Y}\overline{\psi_{L}^{Y}}\Phi_{2}Y_{R}
 +y^{D}\overline{\psi_{L}^{Y}}\widetilde{\Phi}_{2}D_{R}
 +h.c.
\label{tlyl}
\end{eqnarray}

\begin{eqnarray}
-\mathcal{L}_{Y}^{L} & = & y_{ij}^{e}\overline{L_{iL}}\widetilde{\Phi}_{1}l_{jR}
+y_{ij}^{\nu}\overline{L_{iL}}\Phi_{1}l_{jR}
\nonumber \\
 & + & x^{A}\overline{\xi_{L}^{Y}}\widetilde{\Phi}_{2}E^{3e}_{R}
 +x^{B}\overline{\xi_{L}^{Y}}\Phi_{2}E^{2e}_{R}
 +h.c.
\label{Ylep}
\end{eqnarray}

\begin{table}
\begin{tabular}{|c|c|c|c|c|c|}
\hline
 & $\Phi_{1}$  & $q_{L}$ &  $L_{L}$ & $S_{1/2}$ \tabularnewline
\hline
\hline
Z$_{4}$  & $w_{2}$ & $w_{2}$ & $w_{2}$ & $w_{2}$ \tabularnewline
\hline
\end{tabular}
\captionsetup{singlelinecheck=off,font=footnotesize}
\caption{\label{zns}Charges of the fields transforming nontrivially  under $Z_{4}$, where $w_{n}\equiv e^{i\frac{\pi n}{2}}$. }
\label{table1}
\end{table} 
\noindent
where, in Eqs.\eqref{tlyl} and \eqref{Ylep} the first-second lines correspond to the SM-Exotic Yukawa Lagrangian for quark and lepton sectors 
respectively. Regarding the scalar sector, the mass eigenstates of the Higgses $H^{0}$ and $h^{0}$ reads,

\begin{equation}
\begin{gathered}
h^{0}=Hs_{\alpha}+hc_{\alpha}\\
H^{0}=Hc_{\alpha}-hs_{\alpha}
\end{gathered}
\label{Higgs}
\end{equation}

Equations \eqref{tlyl}, \eqref{Ylep} and \eqref{Higgs} give the interaction terms between the exotic fermions and the Higgses, 
where in the mass eigenstates reads,

\begin{equation}
L_{\mathcal{QH}}=\sum_{i}\frac{\sqrt{2}m_{qi}}{vs_{\beta}}\bar{q}_{iL}q_{iR}(Hs_{\alpha}+hc_{\alpha})+
\sum_{\eta}\frac{\sqrt{2}M_{\eta Q}}{vs_{\beta}}\bar{Q}_{\eta L}Q_{\eta R}(Hc_{\alpha}-hs_{\alpha})
\label{LQH}
\end{equation}

in \eqref{LQH} the standard relations between the vevs $tg\beta=\frac{v_{2}}{v_{1}}$ have been used. The exact decoupling limit is 
reached taking $\alpha=0$, in which the contributions to the SM-Higgs production coming from exotic quarks ($h\bar{Q}Q$) are absent.
The Lagrangian $L_{\mathcal{QH}}$ in this limit takes the form,

\begin{equation}
L_{\mathcal{QH}}=\sum_{i}\frac{\sqrt{2}m_{qi}}{vs_{\beta}}\bar{q}_{iL}q_{iR}h+
\sum_{\eta}\frac{\sqrt{2}M_{Q\eta}}{vc_{\beta}}\bar{Q}_{\eta L}Q_{\eta R}H
\label{HiggsQq}
\end{equation}
where $i$ run over the SM-fermions and $\eta$ on the new fermions.
Equation \eqref{HiggsQq} shows that the fermion masses are strongly dependent of the $\beta$ angle. It imposes upper bounds on large masses
through demand perturvativity of the Yukawa couplings $y^2\lesssim 4\pi$. The still allowed values of $\beta$ depends on the type of 
two Higgs doublet model (I,II,III IV) that is used as scalar sector. In addition, $\beta$ 
has been strongly suppressed by electroweak measurements, in which, the so-called alignment limit suggest that $cos(\beta)\leq 0.4$ in order 
to keep all the SM predictions into the $95\%$ CL \cite{Dumont:2014wha}. The above considerations impose an upper limit on the masses of the exotics 
fermions in $M_{Q}\sim 300$ GeV, which leads to expect that the exotic fermions into this new framework will be manifests at the scale of the 
EWSB. Other experimental 
limits have already been imposed in many works for the $Y_{-4/3}$, $D_{-1/3}$ exotic quarks masses assuming a completely different decay 
chains and branching ratios to the current work \cite{D0-exquark-2007}\cite{exotic_charge}. 

The idea to look for new physics at the EWSB scales motivate the construction of models containing new decay channels so far unexplored 
for the exotic fermions. The interaction Lagrangian in equation \eqref{LQQ} allows the decay of all the exotic fermions always 
through the interaction with a SLQ, 
\begin{equation}
\begin{gathered}
-\mathcal{L}_{LQ} = \lambda_{d}\overline{d}_{R}S_{1/2}^{T}\epsilon L_{L}+\lambda_{D}\overline{D}_{R}S_{1/2}^{T}\epsilon L_{L}+\lambda_{q}\overline{\nu^{c}_{R}}u_{R}\tilde{S_{u}}+
\lambda_{A}\overline{E^{2e,c}_{R}}Y_{R}\tilde{S_{u}}+h.c.
\end{gathered}
\label{LQQ}
\end{equation}

where, $\epsilon$ represents two dimensional antisymmetric matrix, $\lambda_{d}$ and $\lambda_{D}$ are matrices in flavor space and
$\lambda_{A}$ and $\lambda_{q}$ are complex numbers. The fields $L$ and $u_{R}$ represent the left-handed SM-lepton doublets and SM-u-type right
handed quarks respectively. Expanding the $SU(2)$ components yields

\begin{equation}
\begin{gathered}
-\mathcal{L}_{LQ} = \lambda_{d}^{\nu}\overline{d}_{R}S_{1/2}^{-1/3}\nu_{L}+\lambda_{d}^{l}\overline{d}_{R}S_{1/2}^{2/3}l_{L}+
\lambda_{D}^{\nu}\overline{D}_{R}S_{1/2}^{-1/3}\nu_{L}+\lambda_{D}^{l}\overline{D}_{R}S_{1/2}^{2/3}l_{L}+\\
\lambda_{q}\overline{\nu^{c}_{R}}u_{R}\tilde{S_{u}}+\lambda_{A}\overline{E^{3e,c}_{R}}Y_{R}\tilde{S_{u}}+h.c.
\end{gathered}
\label{LQQ1}
\end{equation}

where, the charge subscript for the fermions is omitted whereas the subscript in the leptoquaks represent the electric charges. From Lagrangian 
\eqref{LQQ1}, the possibilities for the $D_{-1/3}$ decay through a SLQ $S^{2/3}_{1/2}$ and $S^{-1/3}_{1/2}$ entails two very different
collider signatures which leads to relax the limits on the masses of the exotic quarks.

\section{Oblique Parameters. \label{section:3}}

In order to suppress possible contributions to the $STU$ parameters due to the new fermionic matter, the scenario is chosen nearly degenerate 
for the doublet of quarks and leptoquarks. For leptons, a narrow split between the masses of the doublet is allowed in order to 
open the $E^{3e}\rightarrow E^{2e} W^{+}$ on-shell decay. Then it is shown that a few scenarios with 
$M_{W}\leq M_{E^{3e}}-M_{E^{2e}}\leq M_{Z}$ GeV are in completet agreement with the last fit for $STU$ \cite{Beringer:1900zz}. Taking
the formulas in Ref~\cite{he-polonsky-su2001}, each pair of fermions
$\left(\psi_{1},\psi_{2}\right)$, with masses $\left(m_{1},m_{2}\right)$,
whose left-handed components form a doublet $\Psi\equiv\left(\psi_{1L}\,\,\psi_{2L}\right)^T\sim(\mathbf{2,\,}\mathcal{Y})$
of hypercharge $\mathcal{Y}$, and their right-handed components are singlets
$\psi_{1R}$, $\psi_{2R}$, gives the following contribution to the
the oblique parameters

\begin{align}
S_{\Psi} & =\frac{N_{C\psi}}{6\pi}\left[1-2\mathcal{Y}\,\mathrm{ln}\frac{x_{1}}{x_{2}}+
\frac{1+8\mathcal{Y}}{20x_{1}}+\frac{1-8\mathcal{Y}}{20x_{2}}\right],\label{ps}\\
T_{\Psi} & =\frac{N_{C}}{8\pi s_{W}^{2}c_{W}^{2}}F\left(x_{1},x_{2}\right),\label{pt}\\
U_{\Psi} & =-\frac{N_{C\psi}}{2\pi}\Bigg\{\frac{x_{1}+x_{2}}{2}
-\frac{\left(x_{1}-x_{2}\right)^{2}}{3}+\left[\frac{\left(x_{1}-x_{2}\right)^{3}}{6}
-\frac{1}{2}\frac{x_{1}^{2}+x_{2}^{2}}{x_{1}-x_{2}}\right]\mathrm{ln}\frac{x_{1}}{x_{2}}\nonumber \\
 & +\frac{x_{1}-1}{6}f\left(x_{1},x_{1}\right)+\frac{x_{2}-1}{6}f\left(x_{2},x_{2}\right)
 +\left[\frac{1}{3}-\frac{x_{1}+x_{2}}{6}-\frac{\left(x_{1}-x_{2}\right)^{2}}{6}\right]f
 \left(x_{1},x_{2}\right)\Bigg\}\label{pu}
\end{align}
in which $N_{C}=3\left(1\right)$ is the color degree of freedom of quarks
(leptons),
\[
F\left(x_{1},x_{2}\right)=\frac{x_{1}+x_{2}}{2}-\frac{x_{1}x_{2}}{x_{1}
-x_{2}}\mathrm{ln}\frac{x_{1}}{x_{2}}
\]
\[
f\left(x_{1},x_{2}\right)=\Bigg\{\begin{array}{c}
-2\sqrt{\Delta}\left[\mathrm{arctan}\frac{x_{1}-x_{2}+1}{\sqrt{\Delta}}
-\mathrm{arctan}\frac{x_{1}-x_{2}-1}{\sqrt{\Delta}}\right]\\
0\qquad\qquad\qquad\qquad\qquad\qquad\qquad\\
\sqrt{-\Delta}\,\mathrm{ln}\frac{x_{1}+x_{2}-1
+\sqrt{-\Delta}}{x_{1}+x_{2}-1-\sqrt{-\Delta}}\qquad\qquad\qquad\qquad\qquad
\end{array}\begin{array}{c}
\left(\Delta>0\right)\\
\left(\Delta=0\right)\\
\left(\Delta<0\right)
\end{array}
\]
with $x_{i}=m_{i}^{2}/M_{Z}^{2}$, and $\Delta=2\left(x_{1}+x_{2}\right)-\left(x_{1}-x_{2}\right)^{2}-1$.

\begin{figure}[!h]
\centering
 \includegraphics[scale=0.43]{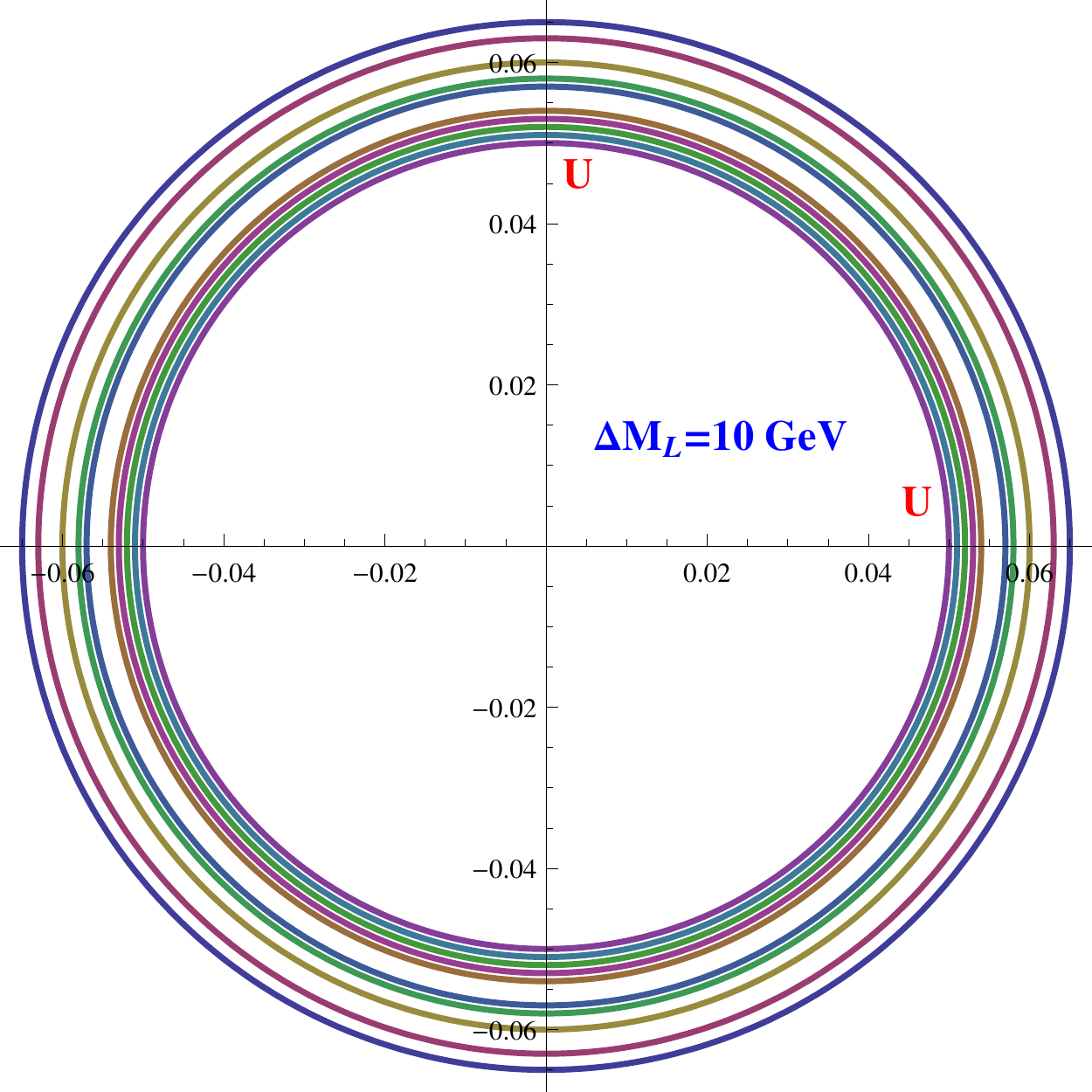} 
  \includegraphics[scale=0.4]{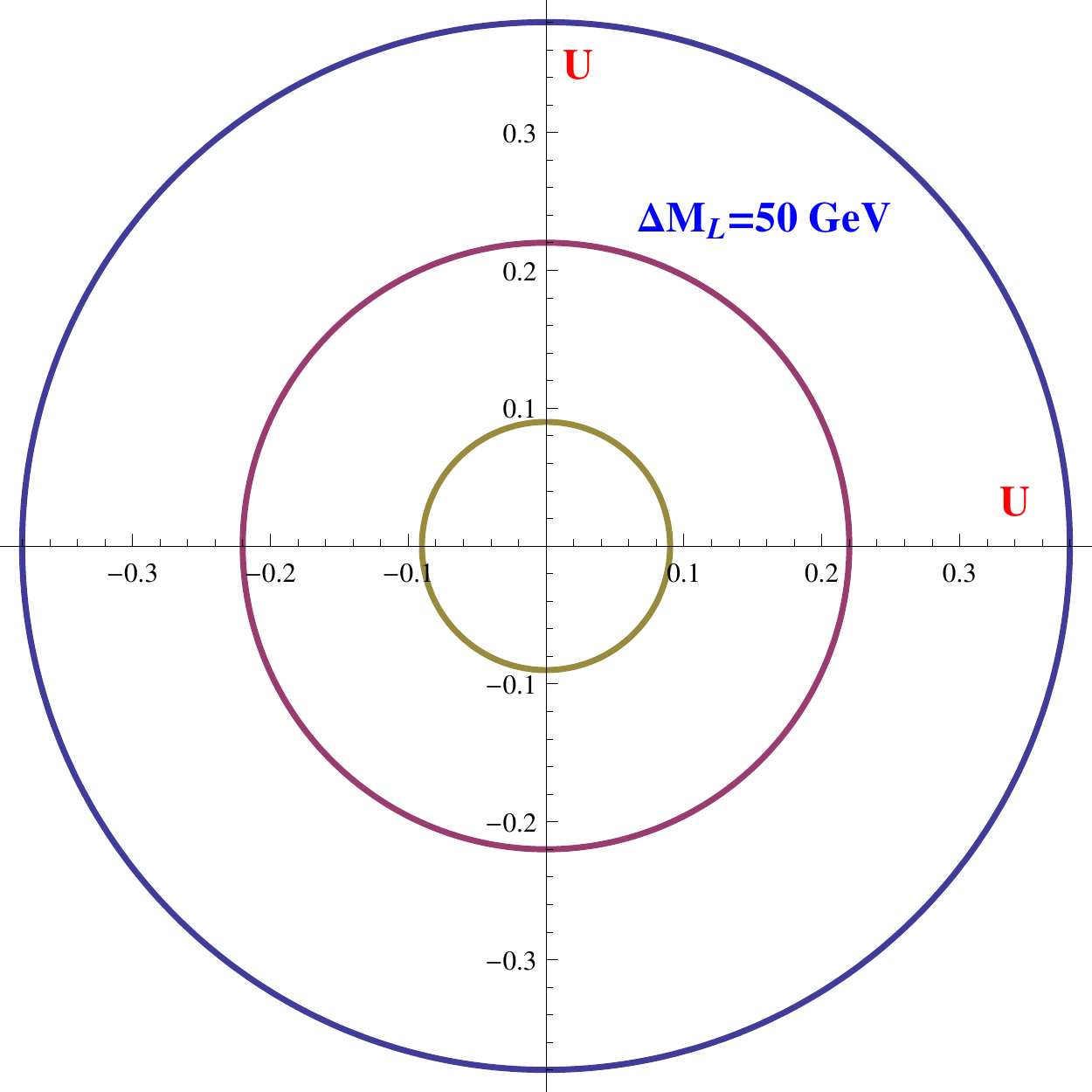}
   \includegraphics[scale=0.4]{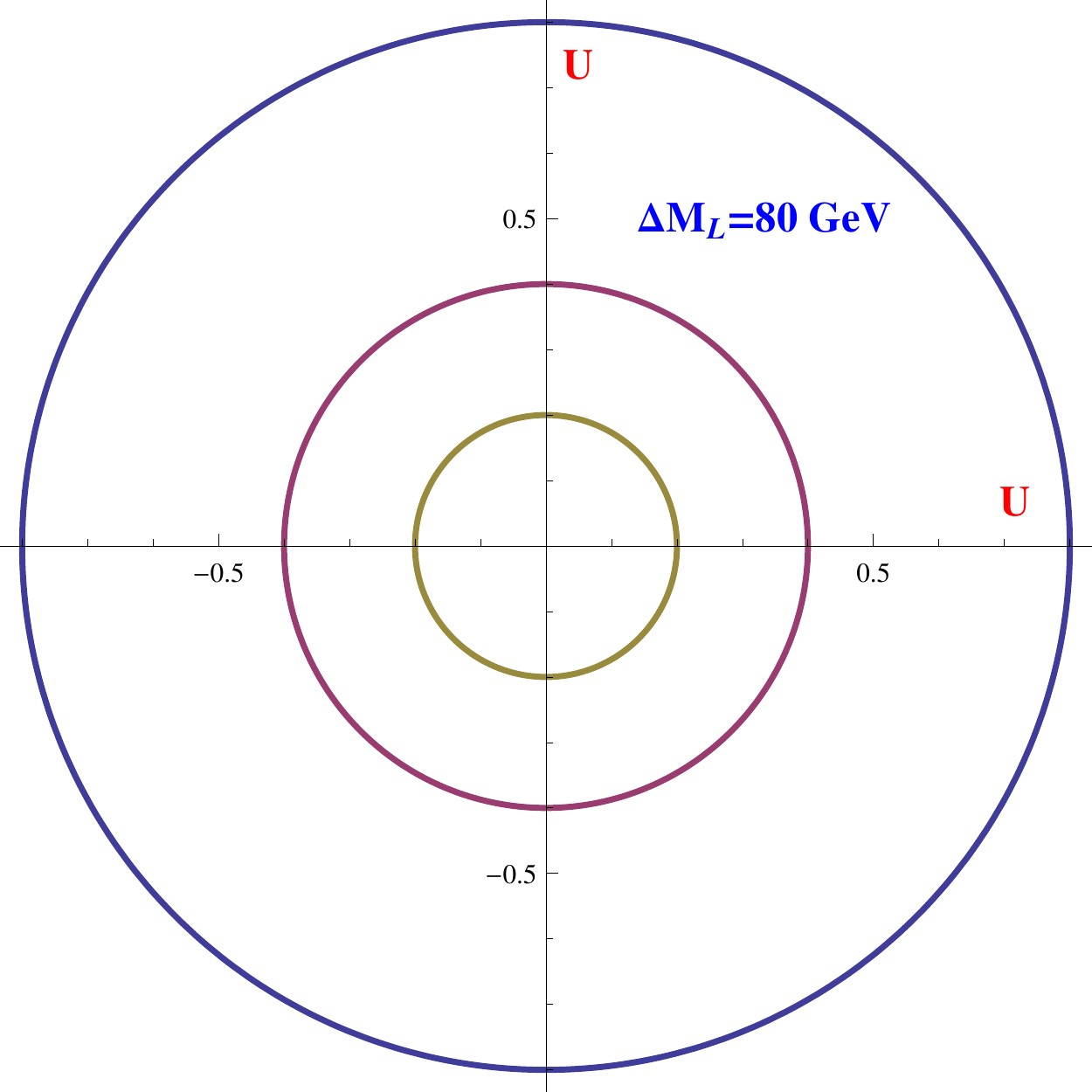}
\captionsetup{singlelinecheck=off,font=footnotesize}
\caption{The radio of each circle represents the value taken by the $U$ parameter when the lepton masses are varied from $M_{E^{3e}}=250$ GeV and $M_{E^{2e}}=240$ GeV for the most outer circle, 
until $M_{E^{3e}}=120$ GeV and $M_{E^{2e}}=110$ GeV for the most inter circle, in which $\Delta M=M_{E^{3e}}-M_{E^{2e}}=10,50\hspace{0.2cm}and\hspace{0.2cm}80$ GeV for the left, middle and right graphs respectively.}

\label{fig:Upara}
\end{figure}

The mass of the new fermionic matter should be greater that $M_{Z}$ in order to avoid $\mathcal{O}(M^{4}_{Z}/M^{4}_{i})$ corrections to the 
above equations. Therefore, from now on, the new matter is assumed to satisfy the $M_{i}>M_{Z}$ condition, in addition,
the mass hierarchy in the doublets makes it always to satisfy $\Delta>0$. Regarding the $S$ parameter, the mass hierarchy between the exotic leptons
implies an offsetting negative contributions from the quarks fields in which is taking the benchmark point with $M_{Y}>M_{D}$. For example, with
$M_{E^{3e}}=250$ GeV, $M_{E^{2e}}=160$ GeV, $M_{Y}=250$ GeV and $M_{D}=240$ GeV gives $S=-0.05$ (in this case $T=0.15$ and $U=0.9$), which show that
the $U$ parameter strongly forbids non-degenerate doublets. In figure \eqref{fig:Upara}, each graph represents the case in which the 
lepton masses are varied in steps of $\Delta M=M_{E^{3e}}-M_{E^{2e}}=10,50\hspace{0.2cm}and\hspace{0.2cm}80$ GeV (left, middle and right)
keeping the quark masses fixed in $M_{Y}=250$ GeV and $M_{D}=240$ GeV. The radio of each circle represents the value
taken by the $U$ parameter when the lepton masses are varied from $M_{E^{3e}}=250$ GeV and $M_{E^{2e}}=240$ GeV for the most external circle, 
until $M_{E^{3e}}=120$ GeV and $M_{E^{2e}}=110$ GeV for the most internal circle. 

From Fig. \eqref{fig:Upara} it can be inferred that all the doublets shall be nearly degenerate in order to keep the electroweak parameters 
under control.

\section{$D_{-1/3}$ Quark Phenomenology. \label{section:4}}
\begin{figure}[!h]
\centering
   \includegraphics[scale=0.8]{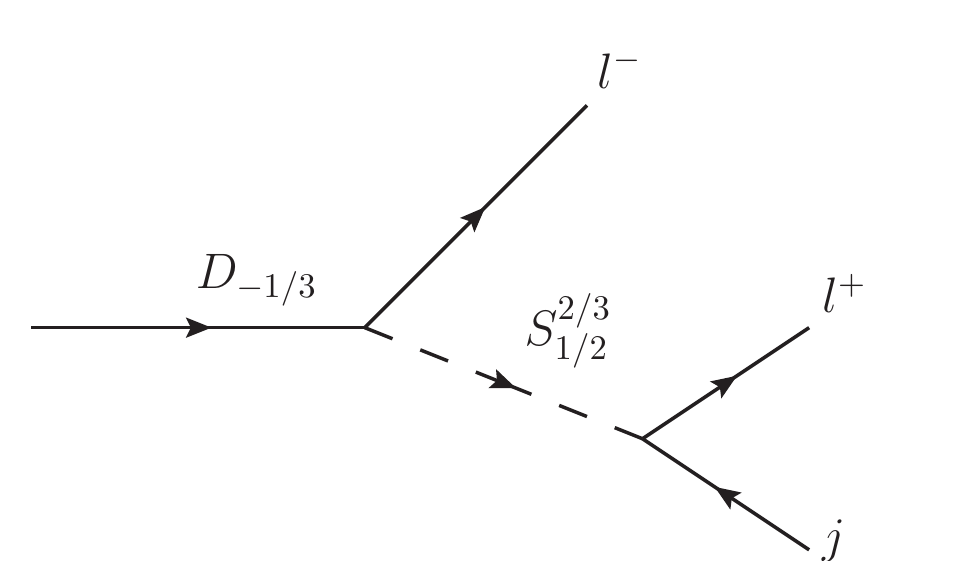}
   \includegraphics[scale=0.8]{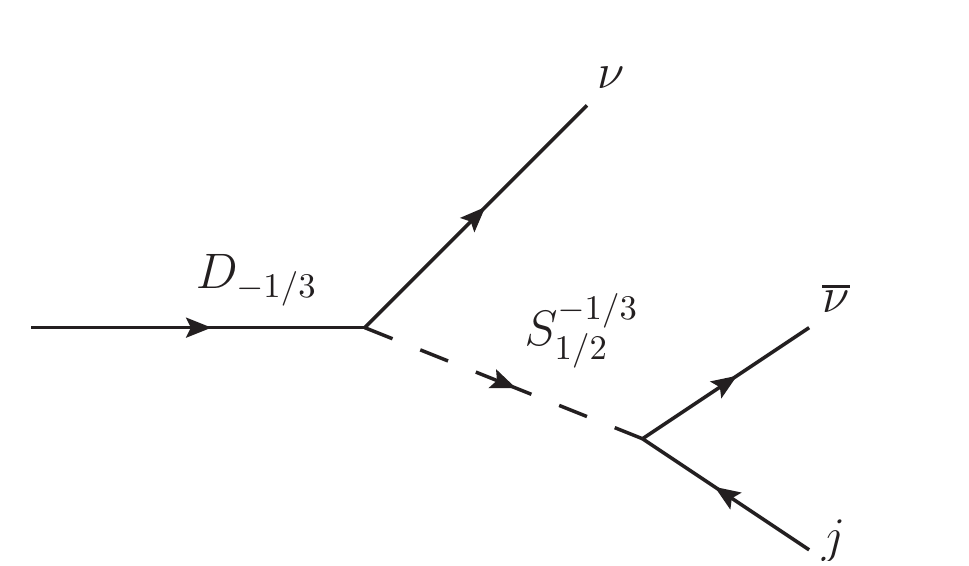}
\captionsetup{singlelinecheck=off,font=footnotesize}  
\caption{Feynman diagrams for the $D_{-1/3}$ decays through a $S^{2/3}_{1/2}$ and $S^{-1/3}_{1/2}$ scalar leptoquaks.}
\label{fig:D1.}
\end{figure}

The most simple decay chain of the non-standard fermions present in this model is for the $D_{-1/3}$ quark. Therefore, in order to test 
the discovery potential of the model or the possible scenarios in which it is potentially excluded, it is introduced the possibles  
signatures for the $D_{-1/3}$ quark at the LHC8. The main production channel is QCD through gluon fusion, followed by the production via 
the decay of $H$ in a pair $D_{-1/3}\bar{D}_{-1/3}$, which, subsequently $D_{-1/3}$ decay in two leptons and a jet through a scalar
leptoquak as mediator Fig. \eqref{fig:D1.}. Limits on scalar leptoquarks masses are less stringent than for the other class of leptoquarks 
\cite{Dorsner:2014axa}. 
Although in this work there are new channels for the scalar leptoquarks which have new couplings with exotic fermions, here, the latest 
fit on their masses has been taken $M_{SLQ}\sim 1$ TeV as a benchmark point. The FeynRules package \cite{FR} has been used to simulate the complete model (SM 
and new interactions) obtaining a UFO \cite{UFO} sampled model containing  all the exotic fermions, heavy scalar and scalar leptoquarks 
as the BSM particles. The necessary coupling terms to be implemented include: the effective $H-hgg$ couplings, $HVV$ couplings, $HQ\bar{Q}$
Yukawa couplings and the leptoquark-quark-lepton interactions $LQ-QL$. 
\begin{table}[!h]
\centering
\begin{tabular}{p{2cm} p{5cm} p{5cm}}
\hline
Parameters &S1:$D_{-1/3}\overline{D}_{-1/3}\rightarrow jj+\notin$& S2:$D_{-1/3}\overline{D}_{-1/3}\rightarrow 2l^{-}2l^{+}+jj$ \\
\hline \hline
$M_{H}$ & 1 TeV & 1 TeV  \\
\hline 
$M_{E^{2e}}$, $M_{E^{3e}}$ & (180, 200) GeV & (180, 200) GeV  \\
\hline 
$M_{D}$, $M_{Y}$ & (220-300, 300) GeV & (220-300, 300) GeV  \\
\hline
$M_{SLQ}$ & 1 TeV & 1 TeV \\
\hline
$g_{SLQ-lQ}$  & 1 (S1A, S1B)& 1 (S2A, S2B)\\
\hline
$g_{SLQ-lq}$ & $10^{-3}$ (S1A, S1B)& $10^{-3}$ (S2A, S2B) \\
\hline
$cos(\beta)$  & 0.35 & 0.35 \\
\hline \hline
\end{tabular}
\captionsetup{singlelinecheck=off,font=footnotesize}
\caption{Representative points in the model parameter space and the relevant mass spectrum used in the analysis.}
\label{tabla:SS}
\end{table}
Let us calculate now the signal cross-section for the completely decay chain for the scenarios S1, S2 in \eqref{tabla:SS}, where $g_{SLQ-lQ}$,
$g_{SLQ-lq}$ and $cos(\beta)$ represent the couplings of the SLQ with one lepton and an exotic quark ($D_{-1/3}$ or $Y_{-4/3}$), the coupling
of the SLQ with one lepton and a SM quark and the value of $cos(\beta)$ in which $\beta=v_{2}/v_{1}$. For each scenario, two complete differents frameworks
(S1A, S1B and S2A, S2B) are taken. The A label means that the SLQ coupled with the first two families of leptons and quarks of the SM, whereas the 
quark $D_{-1/3}$ coupled with the three families of SM-leptons and one SLQ. The B label means that the SLQ coupled just with the third family of 
quarks and leptons of the SM, whereas the quark $D_{-1/3}$ coupled with only the third family of SM-leptons and one SLQ.

\begin{figure}[!h]
\centering
   \includegraphics[scale=0.6]{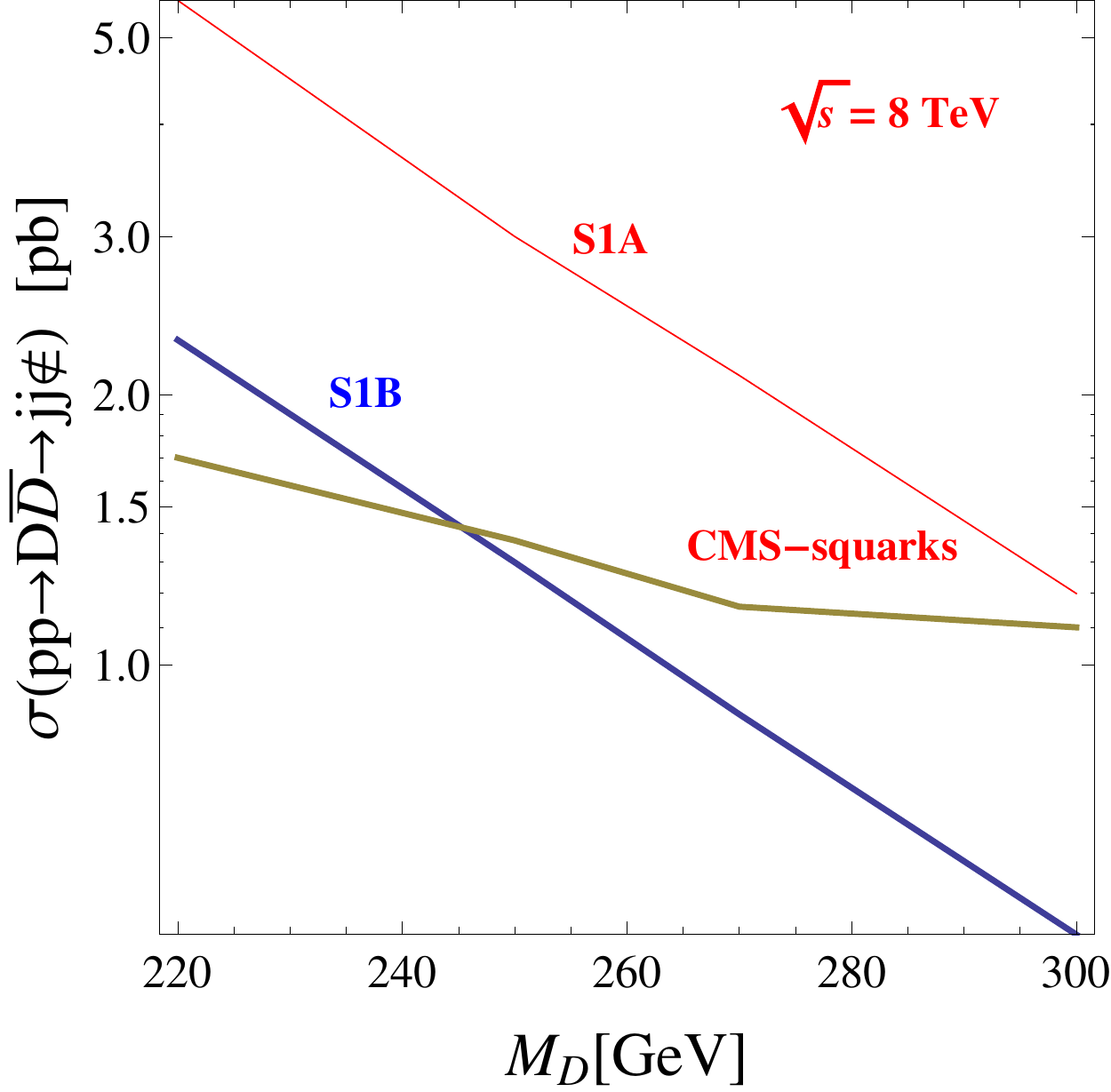}
   \includegraphics[scale=0.62]{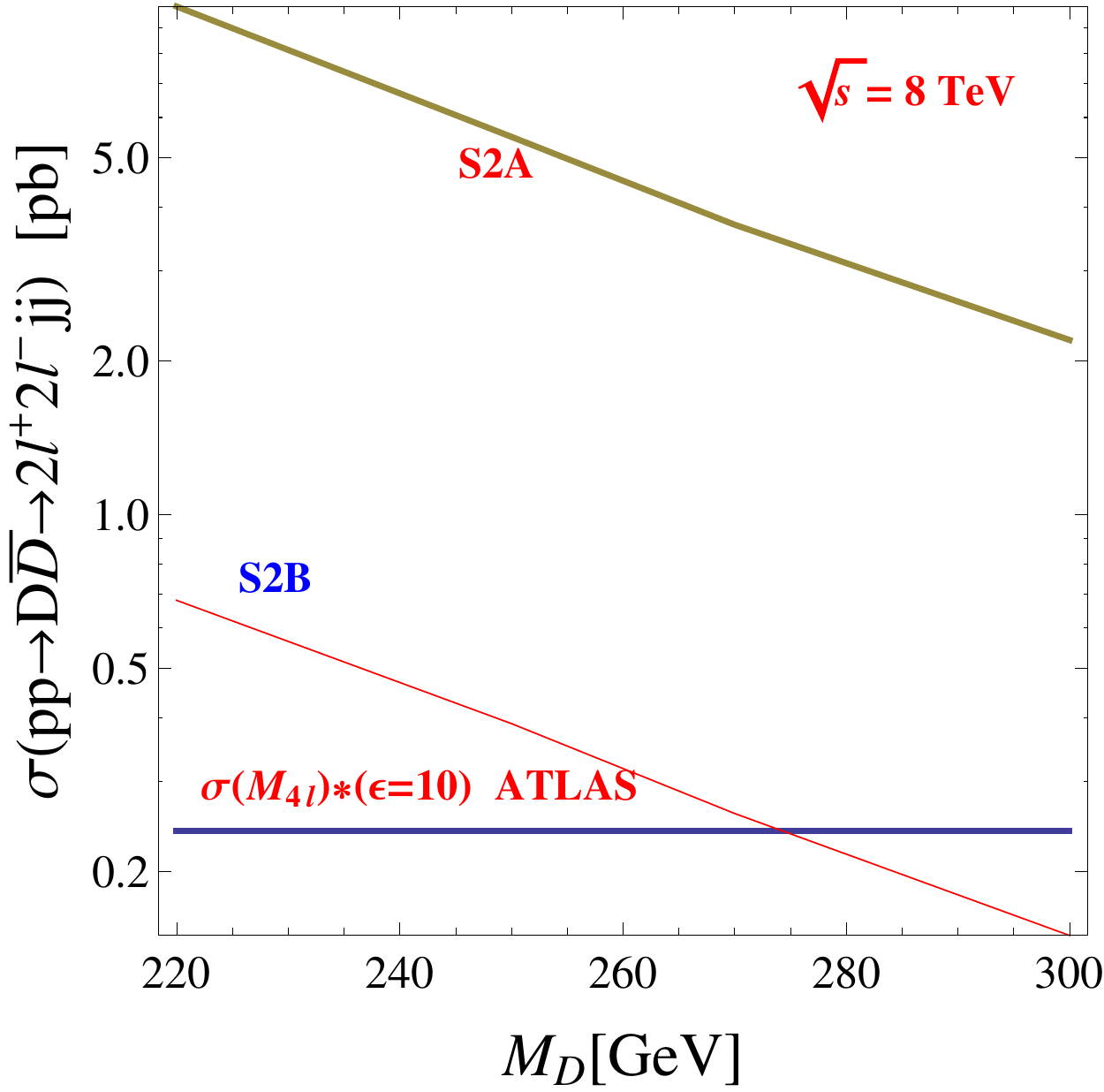}
\captionsetup{singlelinecheck=off,font=footnotesize}   
\caption{Cross-section of the $D_{-1/3}\overline{D}_{-1/3}$ pair production and decay and direct (left) and indirect (right) constraints for 
each decay channel at the LHC8.}
\label{fig:ppD}
\end{figure}
In spite of not having direct limits for both of the possible final states signals of $D_{-1/3}$, searches in SUSY brings us upper limits for 
the squarks cross-sections when they decay in jets plus missing energy, in which, the squarks decays have almost the same final
state of $D_{-1/3}$ when it decays through $S_{1/2}^{-1/3}$. On the other hand, searches containig four leptons serves 
as an idea of the size of the cross-section for the $D_{-1/3}$ decay through $S_{1/2}^{2/3}$ in relation to that observed so far.

\subsection{$D_{-1/3}\overline{D}_{-1/3}\rightarrow jj+\notin$ Channel.}
In \cite{CMS:zxa}, the last upper bounds on cross-sections have been imposed for pair-produced squarks, each decaying to a quark and a neutralino.	 
The topology of the decays are equal for squarks and $D_{-1/3}$ when it decays through a $S_{1/2}^{-1/3}$ leptoquak enabling it impose upper
limits on the $pp\rightarrow D_{-1/3}\overline{D}_{-1/3}\rightarrow jj+\notin$ cross-sections. In Fig. \eqref{fig:ppD} (left) is the 
cross-section for the scenarios S1A and S1B along with the fifth-order polynomial fit of the cross-section for pair-produced squarks decaying in quarks
and a massless neutralino in the LHC at 8 TeV. From it, the scenario S1A is completely exclued, whereas the scenario S1B allows a narrow window of masses 
(240-300) GeV still in agreement with the experimental limits. 

\subsection{$D_{-1/3}\overline{D}_{-1/3}\rightarrow 2l^{-}2l^{+}+jj$ Channel.}
Unlike the previous case, the decay through a $S_{1/2}^{2/3}$ leptoquak has not associated direct limits. 
Nevertheless, the four leptons as a final state lets consider searches in four leptons as indirect constraints on the 
size of $D_{-1/3}\overline{D}_{-1/3}\rightarrow 2l^{-}2l^{+}+jj$ cross-section. Here, we calculate the sum of all the bins of the four
lepton mass invariant graph, figure 13 in \cite{Aad:2014eva} from 200 GeV to 600 GeV divided by the luminosity $L=20.3 fb^{-1}$. An arbitrary factor of
$\epsilon=10$ has been introduced to recompense the signal loss due to the cuts and then obtaining the $\sigma(M_{4l})*(\epsilon=10)$ fit. 
In Fig. \eqref{fig:ppD} (right) is the cross-section for the scenarios S2A and S2B along with the $\sigma(M_{4l})*(\epsilon=10)$ fit. From it, 
we can roughly say that the scenario S2B is strongly excluded whereas the scenario S2B allows a narrow window of masses (270-300) GeV still in 
agreement with the setting made here.

\section{Conclusions}

In this work, the analysis of signatures at the LHC at 8 TeV of a complete renormalizable chiral fermion extension of the SM, which, embeds 
quarks and leptons with exotic charges and scalar leptoquaks has been considered. The exotic fermions mix with scalar leptoquaks via Yukawa couplings with ordinary
leptons keeping the usual couplings of the leptoquaks with SM fermions, even though new coulings with exotic fermions are introduced for the scalar 
leptoquars considered here. Due to the chirality of the new fermions, an extra scalar doublet has been introduced 
in order to give the mass of them through the Higgs mechanism as done in \cite{Alves:2013dga}, where, a benchmark point in complete agreement with the 
last fit for the 2HDM was taken into account.\\
The analysis are based on the comparison of the size of final states cross-sections for the two possibles decays of the quark $D_{-1/3}$ with direct 
and indirect constraints. The direct constraints are coming from supersymmetry searches of squarks which have the same final state signature of
$D_{-1/3}$ decaying in $S_{1/2}^{-1/3}$ and a ordinary lepton. On the other hand, indirect constraints are coming from the complete background for 
the four lepton final state, which, can bring us an idea of how large the model signatures are from the experimental data at 8 TeV. Two benchmark
points are analysed for each scenario, in one (SB) the SLQ couples with the two first families of SM fermions whereas $D_{-1/3}$ couples with the
third SM leptons plus a SLQ, the other sub scenario (SA) is when the SLQ only couples with the third family of SM fermions and $D_{-1/3}$ with 
a lepton of the third SM family and a SLQ.\\
Concerning to the  $D_{-1/3}\overline{D}_{-1/3}\rightarrow jj+\notin$ channel, constraints coming from supersymmetry searches exclude completely the 
scenario S1A, whereas allow a narrow window of masses into the experimental limits (240-300) GeV. For the 
$D_{-1/3}\overline{D}_{-1/3}\rightarrow 2l^{-}2l^{+}+jj$ channel, the fit made here shows that the cross-section 
is too large for the scenario S2A. Nevertheless, the scenario S2B left a narrow window of masses (280-300) GeV still in agreement with 
the indirect constraint. Both scenarios can be easily tested by the new run of the LHC thus leading to provide new channels to search exotic
physics at the LHC energies.

\section{Acknowledgments}
I am grateful to Elmer Barreto for valuable suggestions and comments. Special thanks go to Alex Gomes Dias for valuable 
discussions regarding the possibility to mix exotic quarks with leptoquarks which is the heart of this paper. I am also very grateful to
Andr$\acute{e}$ Lessa for providing me useful ideas and comments during the course of 
this work. I thank Grupo de Simulac$\tilde{a}$o e Modelagem and Universidade Federal do ABC for providing computing 
resources and spaces for this work. 

\end{document}